\documentclass[preprint,12pt]{elsarticle}

\usepackage{geometry}
\usepackage{amssymb}
\usepackage{amsmath}
\usepackage{amsfonts}
\usepackage{amssymb}
\usepackage[version=4]{mhchem}
\usepackage{stmaryrd}

\usepackage[colorlinks,
            linkcolor=blue,
            anchorcolor=blue,
            citecolor=blue]{hyperref}

\usepackage{booktabs}
\usepackage{multirow}

\journal{}

\begin{document}

\begin{frontmatter}



\title{Correction to the quantum relation of photons in the Doppler effect based  on a special Lorentz violation model}


\author[inst1]{Jinwen Hu}
\ead{200731890025@whu.edu.cn}

\author[inst1]{Huan Hu}

\affiliation[inst1]{organization={Department of physics and technology},
            addressline={Wuhan university},
            city={Wuhan},
            postcode={430060},
            country={China}}

\begin{abstract}
The possibility of the breaking of Lorentz symmetry has been discussed in many models of quantum gravity. In this paper we follow the Lorentz violation model in Ref. \cite{b14} (i.e., our previous work) to discuss the Doppler frequency shift of photons and the Compton scattering process between photons and electrons, pointing out that following the idea in Ref. \cite{b14} we have to modify the usual quantum relation of photons in the Doppler effect. But due to the current limited information and knowledge, we could not yet determine the specific expression for the correction coefficient in the modified quantum relation of photons. However, the phenomenon called spontaneous radiation in a cyclotron maser give us an opportunity to see what the expression for this correction coefficient might look like. Therefor, under some necessary constraints, we construct a very concise expression for this correction coefficient through the discussion of different cases. And then we use this expression to analyze the wavelength of radiation in the cyclotron maser, which tends to a limited value at $\boldsymbol{v} \rightarrow \boldsymbol{c}$, rather than to 0 as predicted by the Lorentz model. And the inverse Compton scattering phenomenon is also discussed and we find that there is a limit to the maximum energy that can be obtained by photons in the collision between extremely relativistic particles and low-energy photons, which conclusion is also very different from that obtained from the Lorentz model, in which the energy that can be obtained by the photon tends to be infinite as the velocity of particle is close to $c$. This paper still follows the purpose in Ref. \cite{b14}  that the energy and momentum of particles (i.e., any particles, including photons) cannot be infinite, otherwise it will make some physical scenarios invalid. When the parameter Q characterizing the degree of deviation from the Lorentz model is equal to 0 , all the results and conclusions in this paper will return to the case as in the Lorentz model, so this paper also provides us with a possible experimental scheme to determine the value of Q in Ref. \cite{b14} , although it still requires extremely high experimental energy.
\end{abstract}




\end{frontmatter}


\section{Introduction}\label{sec1}
The study of quantum gravity has always been a frontier and hot topic in fundamental physics, the main reason is that people find that the singularity problem and information loss problem have been arisen in the study of black hole and also find that general relativity and quantum theory are incompatible under the current framework \cite{b1,b2,b3,b4}. The starting point for many quantum gravity models is the belief that Lorentz symmetry may be violated at ultra-high energy scale, such as near the Planck energy scale. The most typical model to hold this view is the double special relativity model (DSR model), which not only considers the speed of light to be a constant, but also introduces another constant called ``minimum length or maximum energy", which leads to a modification to the energy-momentum dispersion relation of particles at extremely high energy scale \cite{b5,b6}. The research on the DSR model can be seen in many literature, and which is believed to be able to solve some challenging problems in the ultra-high energy field \cite{b7,b8,b9,b10,b11,b12}.

However, the DSR model has an obvious disadvantage that it is not concise enough. The equation describing the dispersion relation of particles usually contains many parameters, for example, in the DSR model, the dispersion relation of particles at Planck energy scale is usually expressed as $E^{2}=\boldsymbol{p}^{2}+m^{2}+\eta L_{p}^{\alpha} \boldsymbol{p}^{2} E^{\alpha}+\mathrm{O}\left(L_{p}^{\alpha+1} E^{\alpha+3}\right)$, where $L_{p}$ denotes the ``Planck length", $\alpha$ is a positive integer and $\eta$ is a real number \cite{b7}. Since the physical significance of these parameters is not clear \cite{b13} , many researchers only take one or two of these parameters to study. For this reason, in Ref. \cite{b14}  (i.e., our previous work) we proposed another possible Lorentz violation model, which contains only one parameter in the dispersion relation equation of particles. More importantly, the Lorentz violation model proposed in Ref. \cite{b14}  naturally returns to the Lorentz model at low and medium energy scale, and when the velocity of particle is close to $c$ the energy of the particle tends to a limited value (which is similar to the DSR model), rather than to be infinite as predicted by the Lorentz model. In order to clarify the purpose and viewpoint of this paper, we can first briefly review the Lorentz violation model proposed in Ref. \cite{b14} .

As one knows in special and general relativity the speed of light occupies a central status, so for most Lorentz violation models, the (local) speed of light is assumed to be variable. In this context, in order to allow it is possible that the (local) speed of light can change between inertial systems, in Ref. \cite{b14}  it starts from a discussion of a general relationship between the speed of light and the velocity of the inertial system, i.e., it assumed that the speed of light observed by an observer moving with a velocity $\boldsymbol{v}$ relative to the light source is $n \boldsymbol{c}$, where $n$ is a variable or invariable dimensionless value. Obviously, due to the fact that many experiments at low or medium energy scale have verified the theory of special relativity and general relativity, there must be some constraints on $n$, that is, $n(\boldsymbol{v}, \boldsymbol{c})=n(-\boldsymbol{v}, \boldsymbol{c})=n(\boldsymbol{v},-\boldsymbol{c})=n(-\boldsymbol{v},-\boldsymbol{c})$ and $n(\boldsymbol{v}$ $=0, \boldsymbol{c})=1$ (The detailed reasons for this constraints can be seen in Ref. \cite{b14} ). Thus based on this assumption the coordinate transformation between two inertial systems moving relative to each other with velocity $\boldsymbol{v}$ is obtained as
\begin{equation}
    \label{eq1}
\left\{\begin{array}{c}
\boldsymbol{x}^{\prime}=\gamma(\boldsymbol{x}-\boldsymbol{v} t) \\
t^{\prime}=\gamma\left(t-\frac{\boldsymbol{v}}{k^2(\boldsymbol{v}, \boldsymbol{c})} \boldsymbol{x}\right)
\end{array}\right.
\end{equation}
where $\gamma(\boldsymbol{v}, \boldsymbol{c})=1 / \sqrt{1-\boldsymbol{v}^2 / k^2}, k(\boldsymbol{v}, \boldsymbol{c})=\sqrt{n \boldsymbol{v} \boldsymbol{c}^2 /(n \boldsymbol{c}-\boldsymbol{c}+\boldsymbol{v})}$. And $n$ satisfies $n(\boldsymbol{v}, \boldsymbol{c})=n(-\boldsymbol{v}, \boldsymbol{c})=n(\boldsymbol{v},-\boldsymbol{c})=n(-\boldsymbol{v},-\boldsymbol{c})$ and $n(\boldsymbol{v}=0, \boldsymbol{c})=1$.

From Eq. (\ref{eq1}) one can obtain that if $\mathrm{d} \boldsymbol{x} / \mathrm{d} t=\boldsymbol{c}$, then $\mathrm{d} \boldsymbol{x}^{\prime} / \mathrm{d} t^{\prime}=n \mathbf{c}$, and if $\mathrm{d} \boldsymbol{x}^{\prime} / \mathrm{d} t^{\prime}=\boldsymbol{c}$, then $\mathrm{d} \boldsymbol{x} / \mathrm{d} t=n \boldsymbol{c}$, which ensures that the two inertial systems are equivalent.

Eq. (\ref{eq1}) is similar in form to the Lorentz transformation (i.e., replacing $c$ with $k$ in Lorentz transformation yields Eq. (\ref{eq1})), and it has been shown in Ref.\cite{b14}  that Maxwell's equations are also covariant based on Eq. (\ref{eq1}). And at the same time, correspondingly, the dispersion relation of particle with rest mass $m_0$, compared to the form as in the Lorentz model, is modified as

\begin{equation}
    \label{eq2}
E^2=\boldsymbol{p}^2 k^2+E_0^2
\end{equation}
where $E_0=m_0 k^2, E=\gamma m_0 k^2$ denotes the particle's total energy, $p=\gamma m_0 v$ denotes the particle's momentum, and $\gamma=\gamma(\boldsymbol{v}, \boldsymbol{c}), k=k(\boldsymbol{v}, \boldsymbol{c})$.

Next, one might easily find a correlation that if the dispersion relation of particles is modified as Eq. (\ref{eq2}), then the usual quantum relation of photons (see Eq. (\ref{eq3})) should also need to be modified accordingly. The reason is that for massless particle, the dispersion relation is $E=\boldsymbol{p k}$ rather than $E=\boldsymbol{p} \boldsymbol{c}$ based on Eq. (\ref{eq2}). Secondly, in Ref. \cite{b14}  it allows the speed of light is variable between inertial systems (especially it allows the observed speed of light tends to 0 when the velocity of light source relative to the observer is close to $\boldsymbol{c}$ ), so the usual quantum relation of photons (i.e., Eq. (\ref{eq3})) will no longer be applicable in this case.
\begin{equation}
    \label{eq3}
\left\{\begin{array}{l}
E=h f \\
\boldsymbol{p}=\frac{h}{\lambda}
\end{array}\right.
\end{equation}
where $h$ is the Planck constant, $f$ and $\lambda$ denotes the frequency and wavelength of light, respectively.

Therefor, this is the issue to be discussed in this paper. However, one may wonder why the quantum theory needs to be modified since it has been so successful so far. The answer is that our current quantum theory is built on the basis of Lorentz symmetry, and if the energy is high enough that the Lorentz symmetry is obviously broken, the quantum theory should be modified. One piece of evidence is the proposal of various models of quantum gravity, which aim to explore the new physics that may emerge when energy is pushed to a very high level.

However, when we try to construct or find such a quantum relation of photon that satisfies Eq. (\ref{eq2}), we found that it is very difficult to determine the specific expression, for that the set of expression satisfying $E=p k$ is large. But fortunately, a physical phenomenon called synchrotron radiation in the cyclotron maser can give us a glimpse of what this expression might look like.

In Ref. \cite{b15,b16} the author used the concept of virtual photons to discuss the phenomenon called spontaneous radiation and synchrotron radiation in a cyclotron maser. By considering the Doppler effect and Compton scattering effect of virtual photons, the author obtained the formulas of spontaneous radiation and synchrotron radiation in the cyclotron maser, which are same as the classical theoretical results. Inspired by the idea in Ref. \cite{b15,b16}, in this paper, we first intend to discuss the constraints of the modified quantum relation of photon, and then try to find a possible expression for the modified quantum relation of photon satisfying Eq. (\ref{eq2}). Further, we intend to discuss the effect of this modified quantum relation of photon on synchrotron radiation in the cyclotron maser. This paper was thus organized as follows. In Section ``Doppler effect of photons", the Doppler effect of photons based on Ref. \cite{b14} is discussed, and in Section ``Compton scattering" the corresponding Compton scattering effect of photons is also discussed. In order to apply both the Doppler effect and Compton scattering effect of photons to one physical phenomenon at the same time, we choose to analyze the phenomena called spontaneous and synchrotron radiation in a cyclotron maser, which is shown in Section ``Spontaneous radiation in cyclotron maser", and at the same time the possible expression for the modified quantum relation of photons is discussed. In Section ``Inverse Compton scattering" the inverse Compton scattering phenomenon is discussed and it shows a significant difference in the predicted results between the model in Ref. \cite{b14} and the Lorentz model. Finally we summary the paper.

\section{Doppler effect of photons}\label{sec2}
\begin{figure}[h]
    \centering
    \includegraphics[width=0.35\linewidth]{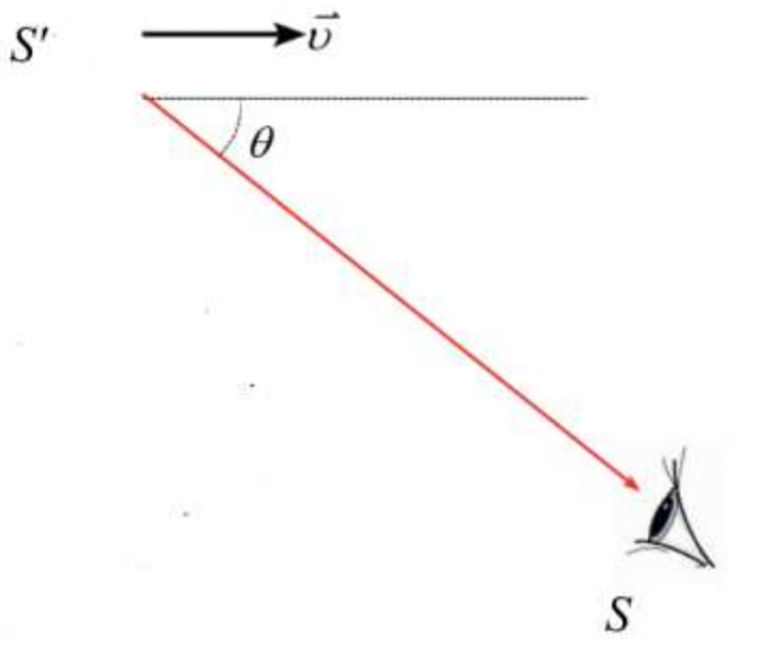}
    \caption{Two inertial systems moving relative to each other in the Doppler effect.}
    \label{fig:01}
\end{figure}

As we know that the frequency of light emitted by the light source is different when measured in an inertial system moving relative to the light source, which is called the Doppler frequency shift effect. As shown in Figure \ref{fig:01} we assume that the inertial system $S^{\prime}$ is moving along the $x\left(x^{\prime}\right)$ axis with a velocity $v$ relative to the inertial system $S$, and the natural frequency and natural wavelength of light emitted by the light source fixed in $S^{\prime}$ is $f^{\prime}$ and $\lambda^{\prime}$, respectively. At the same time we assume the angle between the direction of light propagation and the $\boldsymbol{x}^{\prime}$-axis is $\theta$. Then based on the conclusion in Ref. \cite{b14}, for the observer in $S$ the clock period of the light source will slow down as
\begin{equation}
    \label{eq4}
T=\frac{T}{\sqrt{1-\beta^2}}=\gamma T^{\prime}
\end{equation}
where $\beta=\boldsymbol{v} / k, T^{\prime}$ and $T$ represent the clock period measured in $S^{\prime}$ and $S$, respectively. And $T^{\prime}=1 / f^{\prime}, \lambda^{\prime} f^{\prime}=c$.

On the other hand, since the light source is moving relative to the observer, the wavelength of light measured in $S$ is
\begin{equation}
    \label{eq5}
    \lambda=n \boldsymbol{c} T-(\boldsymbol{v} \cos \theta)T
\end{equation}
where $n=n(\boldsymbol{v} \cos \theta, \boldsymbol{c})$.

From Eqs. (\ref{eq4}) and (\ref{eq5}), we can obtain
\begin{equation}
    \label{eq6}
\left\{\begin{array}{c}
f=\frac{n \boldsymbol{c}}{\lambda}=\frac{\sqrt{1-\beta^2}}{1-\frac{\boldsymbol{v} \cos \theta}{n \boldsymbol{c}}} f^{\prime} \\
\lambda=n\left(1-\frac{\boldsymbol{v} \cos \theta}{n \boldsymbol{c}}\right) \frac{\lambda^{\prime}}{\sqrt{1-\beta^2}}
\end{array}\right.
\end{equation}
where $f$ and $\lambda$ is the frequency and wavelength of light measured in $S$, respectively.

Eq. (\ref{eq6}) shows the corresponding Doppler effect formula based on the conclusion from Ref. \cite{b14}, and it can be seen that when $n \equiv 1$, Eq. (\ref{eq6}) returns to the usual case as in the Lorentz model. Here one may wonder why the mathematical forms of Eqs. (\ref{eq1}) and (\ref{eq2}) are similar to that in the Lorentz model (i.e., replacing $c$ with $k$ in the Lorentz model yields Eqs. (\ref{eq1}) and (\ref{eq2})), but Eq. (\ref{eq6}) has no similarity with the relativistic Doppler frequency shift formula in the Lorentz model. The main reason is that, in Eq. (\ref{eq1}), the speed, wavelength, and frequency of light are all variable, while in the Lorentz model, only the wavelength and frequency of light are variable. Therefor, if the mathematical formulas in Eq. (\ref{eq6}) is still similar in form to that in the Lorentz model, then $k$ will eventually be derived as a constant (note that in Eq. (\ref{eq1}) $k$ is a function of $\boldsymbol{v}$ ), which is obviously contradictory to Eq. (\ref{eq1}).

Based on Eq. (\ref{eq6}), next we discuss a special case where $\theta=0$ (if $\theta=$ $\pi / 2$, then $n(0, c)=1$, Eq. (\ref{eq6}) returns to the case as in the Lorentz model), in which case one can obtain that
\begin{equation}    \label{eq7}
\left\{\begin{array}{c}
f=\frac{n \boldsymbol{c}}{\lambda}=\frac{\sqrt{1-\beta^2}}{1-\frac{\boldsymbol{v}}{n \boldsymbol{c}}} f^{\prime} \\
\lambda= n\left(1-\frac{\boldsymbol{v}}{n \boldsymbol{c}}\right) \frac{\lambda^{\prime}}{\sqrt{1-\beta^2}}
\end{array}\right.
\end{equation}

Here it should be noted that  $n=n(\boldsymbol{v}, \boldsymbol{c})$ in Eq. (\ref{eq7}) due to $\boldsymbol{v} \cos \theta=\boldsymbol{v}$.

Based on Eq. (\ref{eq7}), one may notice that when $\boldsymbol{v}=n \boldsymbol{c}, \lambda=0$ and $f$ tends to be infinite, which is similar to the property of shock wave in the acoustic Doppler effect. Similarly, from Ref. \cite{b14} we can obtain the condition for light to generate the shock wave, that is, according to Ref. \cite{b14}, when $\boldsymbol{v}$ approaches $\boldsymbol{c}$, there is
\begin{equation}
    \label{eq8}
\begin{aligned}
n & =\frac{1}{1-Q}\left[1-Q^{1-\left(\frac{v}{c}\right)^2}\right]=\frac{1}{1-Q}\left[1-Q^{\left(1+\frac{v}{c}\right)\left(1-\frac{v}{c}\right)}\right] \\
& \approx \frac{1}{1-Q}\left[1-Q^{2\left(1-\frac{v}{c}\right)}\right] \approx \frac{2 \boldsymbol{\ln} Q}{Q-1}\left(1-\frac{v}{c}\right)
\end{aligned}
\end{equation}
where $Q$ is a constant and its value needs to be determined by the experiment. As mentioned in Ref. \cite{b14}, we do not yet know the specific value of $Q$ because the current experimental energy is not high enough, but it can be known from a large number of existing experimental results that $Q \approx 0$.

\begin{figure}[h]
    \centering
    \includegraphics[width=0.5\linewidth]{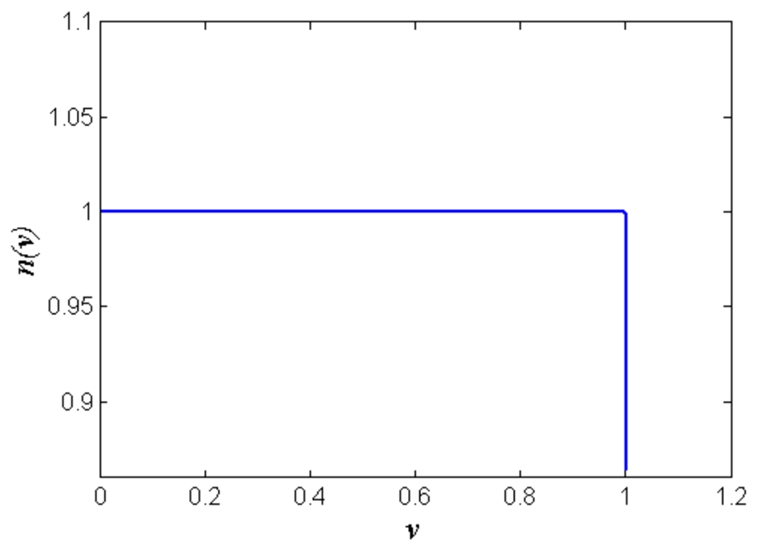}
    \caption{The curve of $n(v) \sim v$ when taking $Q=(1 / e)^{10^6}($ set $c=1)$. Since here it shows the global picture of the curve, a transition arc at the corner is too small to show \cite{b14}. From here or Eq. (\ref{eq8}) it can be seen that when $v=1, n=0$.}
    \label{fig:02}
\end{figure}

The reason why $n$ is chosen as the form in Eq. (\ref{eq8}) is that the function of Eq. (\ref{eq8}) can be well consistent with various experimental results conducted at low or medium energy scale at present, which has been explained in Ref. \cite{b14}. And also importantly, Eq. (\ref{eq8}) is very concise for it has only one parameter $Q$. Here let us re-draw the curve of $n-\boldsymbol{v}$ as in Figure  \ref{fig:02}.

$n$ can be viewed as the degree to which the speed of light observed by the observer in an inertial systems moving relative to the light source deviates from $\boldsymbol{c}$. As can be seen from Figure \ref{fig:02} that $n$ is almost equal to 1 over a large range of $\boldsymbol{v}$. And it is because of this property of $n$ that the model in Ref. \cite{b14} does not violate a large number of existing experimental results conducted at low or medium energy scale. When $Q=0$, $n \equiv 1$, and correspondingly, Eqs. (\ref{eq1}) and (\ref{eq2}) return to the case as in the Lorentz model.

As is well known, when the velocity of the wave source is equal to the speed of wave propagation, the wavefront in the direction of wave source moves cannot propagate outward, but closely adheres to the wave source. This phenomenon is called ``shock wave", which usually occurs when the velocity of an object in a medium changes from subsonic to supersonic, and it is called acoustic shock wave.

Similarly, we can also obtain the condition for light to generate the shock wave. That is, for $\boldsymbol{v}=n \boldsymbol{c}$, there is
\begin{equation}
    \label{eq9}
\boldsymbol{v}_{\text {wave shock }} \approx \frac{1}{\frac{Q-1}{2 \ln Q}+1} \boldsymbol{c} < \boldsymbol{c}
\end{equation}
where $\boldsymbol{v}_{\text {wave shock }}$ is the solution for $\boldsymbol{v}=n \boldsymbol{c}$ and Eq. (\ref{eq9}) is just the condition for generating a shock wave of light.

\begin{figure}[h]
    \centering
    \includegraphics[width=0.35\linewidth]{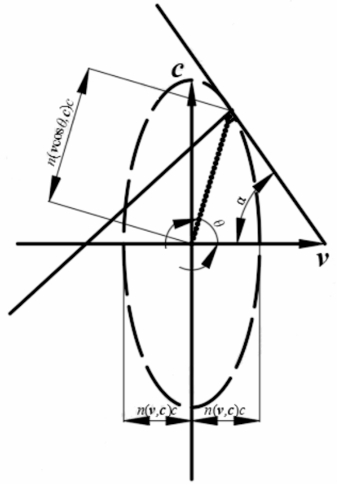}
    \caption{Geometric diagram of the Mach cone of light.}
    \label{fig:03}
\end{figure}

Similar to the phenomenon in supersonic, we can also define the Mach cone of light in the same ways as the Mach cone of acoustic, which is shown in Figure \ref{fig:03}, and where $\alpha$ is the angle of cone.

In Figure \ref{fig:03} the spatial distribution of the speed of light along the center of the light source satisfies the following curve equation $\rho=n(v \cos \theta, c) c$ in polar coordinates. When $v>v_{\text {wave shock, }}$, a reverse timing sequence will occur, which just similar to the phenomenon in supersonic. But as we know, the timing sequence can be reversed in independent events without violating the law of causality (but not in dependent events) \cite{b17}.

\section{Compton scattering}\label{sec3}
We know from the Compton scattering process that light can be regarded as a particle (i.e., photon), in special relativity, the energy and momentum of photons can be expressed by Eq. (\ref{eq3}). However, if substituting Eq. (\ref{eq7}) into Eq. (\ref{eq3}), we can obtain
\begin{equation}
    \label{eq10}
\frac{E}{\boldsymbol{p}}=n \boldsymbol{c}
\end{equation}

Eq. (\ref{eq10}) does not agree formally with Eq. (\ref{eq2}), for that based on Eq. (\ref{eq2}), for massless particles there is $E=\boldsymbol{p} k$. So we will have to modify the quantum relation of photons if we follow the idea in Ref. \cite{b14}. Here we assume the modified quantum relation as
\begin{equation}
    \label{eq11}
\left\{\begin{array}{l}
E=a_1 h f \\
\boldsymbol{p}=a_2 \frac{h}{\lambda}
\end{array}\right.
\end{equation}

Then based on $E=\boldsymbol{p} k$ it has $a_1 / a_2=k / n$. But unfortunately, there seems to be no information or knowledge available to help us specify the expression for $a_1$ and $a_2$.

\begin{figure}[h]
    \centering
    \includegraphics[width=0.5\linewidth]{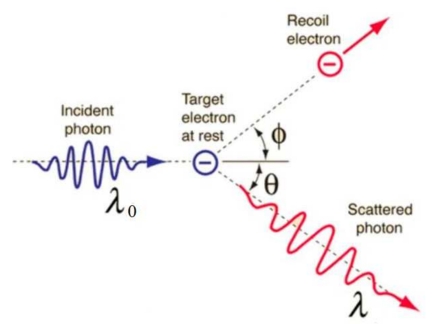}
    \caption{ The diagram of Compton scattering. It shows the elastic collision process between two particles (i.e. a moving photon and a resting electron), which satisfies the energy and momentum conservation.}
    \label{fig:04}
\end{figure}

Next we attempt to discuss the Compton scattering process between photons and electrons based on Eq. (\ref{eq11}), and the corresponding diagram can be seen in Figure \ref{fig:04}. Here for simplicity, we assume the electrons are stationary relative to the laboratory before the collision occurs.

According to the diagram of interaction between photon and electron in Figure \ref{fig:04}, we can set out the following energy and momentum conservation equations
\begin{equation}\label{eq12}
\left\{\begin{aligned}
& a_1 h f_0+m_0 \boldsymbol{c}^2=a_1^{\prime} h f+m k^2 \\
& a_2 \frac{h}{\lambda_0}=a_2^{\prime} \frac{h}{\lambda} \cos \theta+m \boldsymbol{v} \cos \varphi \\
& a_2^{\prime} \frac{h}{\lambda} \sin \theta=m \boldsymbol{v} \sin \varphi \\
& \frac{a_1}{a_2}=\frac{k}{n} \\
& \frac{a_1^{\prime}}{a_2^{\prime}}=\frac{k^{\prime}}{n^{\prime}}
\end{aligned}\right.
\end{equation}

Note that in Eq. (\ref{eq12}) $a_1, a_2, k, n, a_{1}^{\prime}, a_{2}^{\prime}, k^{\prime}, n^{\prime}$ are both functions of $\boldsymbol{v}$ and $\boldsymbol{c}$. And Eq. (\ref{eq12}) can be approximately simplified, that is, when $\boldsymbol{v} \ll \boldsymbol{c}$, $k \approx c$, and when $\boldsymbol{v} \approx \boldsymbol{c}$, there is

\begin{equation}\label{eq13}
\lim _{\boldsymbol{v} \rightarrow \boldsymbol{c}} k=\lim _{\boldsymbol{v} \rightarrow \boldsymbol{c}} \sqrt{\frac{n \boldsymbol{v} \boldsymbol{c}^2}{n \boldsymbol{c}-\boldsymbol{c}+\boldsymbol{v}}}=\lim _{\boldsymbol{v} \rightarrow \boldsymbol{c}} \sqrt{\frac{\boldsymbol{v} \boldsymbol{c}^2}{\boldsymbol{c - \frac { c - v } { n }}}}=\sqrt{\frac{\boldsymbol{c}^2}{1-\frac{Q-1}{2 \ln Q}}}
\end{equation}

Since $Q \approx 0$, then based on Eq. (\ref{eq13}), when $\boldsymbol{v} \approx \boldsymbol{c}$, there is $k \approx c$. To summarize, for $\boldsymbol{v} \in[0, \boldsymbol{c})$, there is $k \approx c$. So we can assume $k \approx k^{\prime} \approx c$ all the time for Eq. (\ref{eq12}) when necessary. Then based on Eq. (\ref{eq12}) we can obtain
\begin{equation}
    \label{eq14}
\frac{\lambda}{a_2^{\prime}}-\frac{\lambda_0}{a_2}=\frac{2 h}{m_0 \boldsymbol{c}^2}(1-\cos \theta) \frac{k\left(\frac{1}{\sqrt{1-\beta^2}}-\frac{c^2}{k^2}\right)}{\frac{2}{\sqrt{1-\beta^2}}-\frac{c^2}{k^2}-\frac{k^2}{c^2}} \approx \frac{h}{m_0 \boldsymbol{c}}(1-\cos \theta)
\end{equation}
where $m_0$ is the rest mass of electron.

Above the formula of Doppler frequency shift and Compton scattering satisfying Eq. (\ref{eq2}) is obtained, but we do not yet know the specific expressions for $a_1$ and $a_2$. What we can do next is put the two formulas into real physical situations to see what the expressions for $a_1$ and $a_2$ might look like. And fortunately, the phenomenon called spontaneous and synchrotron radiation in the cyclotron maser just provide such a physical application.

\section{Spontaneous radiation in cyclotron maser}\label{sec04}
Cyclotron maser is a kind of device with high peak power and high average power in millimeter band and sub-millimeter band. After decades of effort, the mechanism, experiment design and numerical simulation subjecting to the cyclotron maser have been analyzed extensively \cite{b18,b19,b20,b21,b22,b23}. For example, in Ref. \cite{b15,b16}, the author found that the Compton scattering mechanism of virtual photons can be used to analyze the spontaneous radiation in a cyclotron maser (In the spontaneous radiation theory of the free electron laser, the periodic static magnetic field in the oscillator is also regarded as virtual photons, which are then converted into real photons by the Compton scattering between virtual photons and relativistic particles \cite{b24}). Inspired by this idea, now we will revisit the discussion on spontaneous radiation in a cyclotron maser using Eqs. (\ref{eq6}) and (\ref{eq14}).

\begin{figure}[h]
    \centering
    \includegraphics[width=0.7\linewidth]{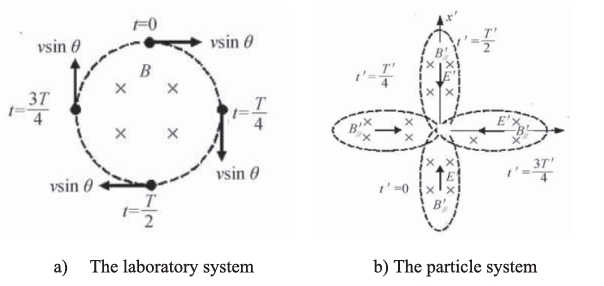}
    \caption{The relative motion between magnetic fields and charged particles \cite{b15}. In a) it shows the particle moves as a circle (or spiral, to be precise) in a resting magnetic field, and in b) it shows the particle is stationary, while the magnetic field moves as a reverse spiral translation around the particle. Due to the symmetry of the inertial system, the two phenomena are equivalent.}
    \label{fig:05}
\end{figure}

In the laboratory system $S$ (i.e., Figure \ref{fig:05}a)), we set the charged particle have a charge of $q$, a rest mass of $m_{0}$, a velocity of $\boldsymbol{v}$ perpendicular to the magnetic field $\boldsymbol{B}$, then in the laboratory system, the spiral motion of the charged particle in a magnetic field has a period of
\begin{equation}\label{eq15}
\begin{aligned}
T_c=\frac{2 \pi R}{\boldsymbol{v}}=\frac{2 \pi m_0}{B q \sqrt{1-\beta^2}}
\end{aligned}
\end{equation}

As shown in Figure \ref{fig:05}b), we construct an inertial systems $S^{\prime}$ that is stationary relative to the particle and the direction of $\boldsymbol{v}$ is parallel to the $\boldsymbol{z}^{\prime}$ axis. Based on the transformation in relativity, there will be both magnetic and electric fields in $S^{\prime}$. And based on the ``moving ruler" effect in relativity, in $S^{\prime}$ the track perimeter of the magnetic field moving as a reverse spiral translation around the charged particle becomes $2 \pi R / \gamma$, then correspondingly, its cycle period is
\begin{equation}
    \label{eq16}
T_{c}^{\prime}=\frac{2 \pi R / \gamma}{\boldsymbol{v}}=T_{c} / \gamma
\end{equation}

\begin{figure}[h]
    \centering
    \includegraphics[width=0.7\linewidth]{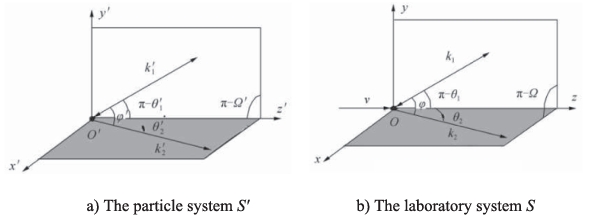}
    \caption{Angle definition in Compton scattering. Due to the symmetry of $S$ and $S^{\prime}$, there is a correlation between the angles defined in the two inertial systems.}
    \label{fig:06}
\end{figure}

According to Ref. \cite{b15}, since the photon in Eq. (\ref{eq16}) cannot be observed in the laboratory system, but can only be observed in $S^{\prime}$ due to the relativistic effect, the photon can be regarded as virtual photon. Next, we will consider the Doppler effect of virtual photon, as both the electric field and magnetic field that vary periodically move with a velocity $-\boldsymbol{v}$ relative to $S^{\prime}$. Now first we define the angle shown in Figure \ref{fig:06}, that is, in the particle system $S^{\prime}, \theta_{1}{ }^{\prime}$ denotes the angle between the direction of virtual photon propagation and the $\boldsymbol{z}^{\prime}$ axis, $\theta_{2}^{\prime}$ denotes the angle between the direction of the scattered photon propagation and the $\boldsymbol{z}^{\prime}$ axis, $\varphi^{\prime}$ denotes the scattering angle. $\Omega^{\prime}$ is the angle between the plane formed by the direction of virtual photon propagation and the $z^{\prime}$ axis and the plane formed by the direction of scattered photon propagation and the $\boldsymbol{z}^{\prime}$ axis, and in the laboratory system $S, \theta_{1}$ denotes the angle between the direction of virtual photon propagation and the $z$ axis, $\theta_{2}$ denotes the angle between the direction of the scattered photon propagation and the $z$ axis, $\varphi$ denotes the scattering angle. $\Omega$ is the angle between the plane formed by the direction of virtual photon propagation and the $z$ axis and the plane formed by the direction of scattered photon propagation and the $z$ axis. Therefor, based on Eq. (\ref{eq6}) we can obtain the wavelength of virtual photons in the particle system $S^{\prime}$ as the source of the virtual photon is moving at a velocity $-\boldsymbol{v}$ relative to the particle system $S^{\prime}$
\begin{equation}
    \label{eq17}
\lambda_1^{\prime}=\frac{\lambda_c^{\prime}}{\sqrt{1-\beta^2}} \boldsymbol{n}\left[1-\frac{\boldsymbol{v}}{n\boldsymbol{c}} \cos \left(\pi-\theta_1^{\prime}\right)\right]
\end{equation}
where $\lambda_c^{\prime}=\boldsymbol{c} T_c{ }^{\prime}=\boldsymbol{c} T_c / \gamma$, and $n=n\left(\boldsymbol{v} \cos \left(\pi-\theta_1{ }^{\prime}\right), \boldsymbol{c}\right)$.

In Compton scattering the virtual photons can be scattered, and radiated out as real photons. And based on Eq. (\ref{eq14}) the wavelength of radiated photons satisfies
\begin{equation}
    \label{eq18}
\frac{\lambda_2^{\prime}}{a_2^{\prime}}-\frac{\lambda_1^{\prime}}{a_2} \approx \frac{h}{m_0}\left(1-\cos \phi^{\prime}\right)
\end{equation}
where $\lambda_2^{\prime}$ denotes the wavelength of radiated photons observed in $S^{\prime}$.

Again, using the Doppler formula in Eq. (\ref{eq6}), we can obtain the wavelength of radiated photons observed by the observer in the laboratory system $S$
\begin{equation}
    \label{eq19}
\lambda_2=\frac{\lambda_2^{\prime}}{\sqrt{1-\beta^2}} n^{\prime}\left(1-\frac{\boldsymbol{v}}{n^{\prime} \boldsymbol{c}} \cos \theta_2\right)
\end{equation}

where $n^{\prime}=n\left(\boldsymbol{v} \cos \theta_2, \boldsymbol{c}\right)$.

Since $Q \approx 0, a_1, a_2$ in Eq. (\ref{eq11}) should satisfy $a_1 \approx 1, a_2 \approx 1$ at low or medium frequency shift. And on the other hand, since the Compton wavelength of the charged particle is generally much smaller than $c T_c$, then we can assume $a_2{ }^{\prime} \approx 1$ in Eq. (\ref{eq18}). So based on Eqs. (\ref{eq14})-(\ref{eq19}), we can obtain
\begin{equation}
    \label{eq20}
\lambda_2=\frac{n^{\prime}-\boldsymbol{v} / \boldsymbol{c} \cos \theta_2}{\sqrt{1-\beta^2}}\left[T_c \frac{n}{a_2}\left(1+\frac{\boldsymbol{v}}{n \boldsymbol{c}} \cos \theta_1^{\prime}\right)+\frac{h}{m_0}\left(1-\cos \phi^{\prime}\right)\right]
\end{equation}

As stated in Ref. \cite{b14} that, the mathematical form of the model proposed in Ref. \cite{b14} is very similar to the Lorentz model, i.e., replacing $c$ in the Lorentz model with $k$ one can obtain the model in Ref. \cite{b14}. Again, based on the same similarity, the formula of aberration of light corresponding to the model in Ref. \cite{b14} can be obtained
\begin{equation}
    \label{eq21}
\left\{\begin{array}{c}
\cos \theta^{\prime}=\frac{\cos \theta-\beta}{1-\beta \cos \theta} \\
\sin \theta^{\prime}=\frac{\sin \theta}{\gamma(1-\beta \cos \theta)}
\end{array}\right.
\end{equation}
where $\theta=\theta_1$ or $\theta=\theta_2$ or $\theta=\Omega$.

On the other hand, according to the geometric definition in Figure \ref{fig:06}, one can obtain the following geometric equations
\begin{equation}
    \label{eq22}
\begin{aligned}
\cos \phi^{\prime} & =\cos \theta_1^{\prime} \cos \theta_2^{\prime}+\sin \theta_1^{\prime} \sin \theta_2^{\prime} \cos \Omega^{\prime} \\
\cos \phi & =\cos \theta_1 \cos \theta_2+\sin \theta_1 \sin \theta_2 \cos \Omega
\end{aligned}
\end{equation}
Substituting Eqs. (\ref{eq21}) and (\ref{eq22}) into Eq. (\ref{eq20}), we can obtain
\begin{equation}
    \label{eq23}
\begin{aligned}
\lambda_2& = \frac{n-\boldsymbol{v} / \boldsymbol{c}cos \theta_2}{\sqrt{1-\beta^2}}\left[T_c \frac{n}{a_2}\left(1+\frac{k}{n \boldsymbol{c}} \frac{\beta \cos \theta_1-\beta^2}{1-\beta \cos \theta_1}\right)\right]\\
&+\frac{h}{m_0} \sqrt{1-\beta^2}(n -\boldsymbol{v} / \boldsymbol{c}cos \theta_2) \frac{1-\cos \phi}{\left(1-\beta \cos \theta_1\right)\left(1-\beta \cos \theta_2\right)}
\end{aligned}
\end{equation}

Eq. (\ref{eq23}) is just the wavelength of spontaneous radiation of a charged particle in a Cyclotron Maser that is observed by the observer in the laboratory system $S$. At low or medium frequency shift, there has $n \approx 1$, $a_2 \approx 1, k \approx 1$, then Eq. (\ref{eq23}) returns to the following equation
\begin{equation}
    \label{eq24}
\lambda_2=T_c \sqrt{1-\beta^2} \frac{1-\beta \cos \theta_2}{1-\beta \cos \theta_1}+\frac{h}{m_0} \sqrt{1-\beta^2} \frac{1-\cos \phi}{1-\beta \cos \theta_1}
\end{equation}
Which corresponds to case as in the Lorentz model.

But here we are more interested in what happens at ultra-high frequency shift. First it reminds us that since it allowed $\boldsymbol{v} \in[0, \boldsymbol{c})$ in Ref. \cite{b14}, when $\boldsymbol{v}=n \boldsymbol{c}$ the energy of the scattered virtual photon in Eq. (\ref{eq12}) maybe infinite due to $\lambda_1{ }^{\prime}=0$ unless we assume
\begin{equation}
    \label{eq25}
\left\{\begin{array}{c}
a_1=\frac{1-\frac{\boldsymbol{V}}{n \boldsymbol{c}}}{1-n \frac{\boldsymbol{V}}{\boldsymbol{c}}} \\
a_2=\frac{n}{k} a_1 \approx n a_1=n \frac{1-\frac{\boldsymbol{V}}{n \boldsymbol{c}}}{1-n \frac{\boldsymbol{V}}{\boldsymbol{c}}}
\end{array}\right.
\end{equation}
where $\boldsymbol{V}=\boldsymbol{v} cos \theta, n=n(\boldsymbol{V}, \boldsymbol{c}), \theta$ is the angle between the direction of $\boldsymbol{v}$ and $\boldsymbol{c}$.

The proposal of Eq. (\ref{eq25}) is based on the following considerations.

Since the expression of $\lambda$ in Eq. (\ref{eq6}) has the term $[1-v \cos \theta /(n c)]$, which leads to $\lambda=0$ under the condition of $\boldsymbol{v}=n \boldsymbol{c}$ and $\theta=0$, and $\lambda=$ 0 will cause the energy of the virtual photon to be infinite, which will make Eq. (\ref{eq14}) invalid. To avoid this, we hope $a_2$ in Eq. (\ref{eq11}) have the term $[1-\nu \cos \theta /(n c)]$ to cancel out this same term in $\lambda$.

On the other hand, we also want to extend Eq. (\ref{eq23}) to the case where $\boldsymbol{v} \rightarrow \boldsymbol{c}$, and in which case, from Ref. \cite{b14} we know that
\begin{equation}
    \label{eq26}
\left\{\begin{aligned}
\lim _{\boldsymbol{v} \rightarrow \boldsymbol{c}} n & =0 \\
\lim _{\boldsymbol{v} \rightarrow \boldsymbol{c}} \frac{1-\boldsymbol{v} / \boldsymbol{c}}{n} & =\frac{Q-1}{2 \ln Q}
\end{aligned}\right.
\end{equation}

Eq. (\ref{eq26}) reminds us that in order to avoid $a_2$ being infinite in the case where $\boldsymbol{v} \rightarrow \boldsymbol{c}$, we can replace $\boldsymbol{v} / \boldsymbol{c}$ with $n \boldsymbol{v} / \boldsymbol{c}$ in the expression for $a_2$, then for $\boldsymbol{v} \rightarrow \boldsymbol{c}$, the expression of $1 /(1-n \boldsymbol{v} / \boldsymbol{c})$ tends to 1 instead of the expression of $1 /(1-\boldsymbol{v} / \boldsymbol{c})$ tends to be infinite, which argument seems a little hard to understand here, we can see how it works in the following. Substituting Eqs. (\ref{eq25}) and (\ref{eq6}) into Eq. (\ref{eq11}), one will obtain
\begin{equation}
    \label{eq27}
\left\{\begin{array}{c}
E=a_1 h f=\frac{\sqrt{1-\beta^2}}{1-\frac{n \boldsymbol{V}}{\boldsymbol{c}}} h f_0 \\
\boldsymbol{p}=a_2 \frac{h}{\lambda}=\frac{E}{k} \approx E
\end{array}\right.
\end{equation}

Based on Eq. (\ref{eq27}), it has

a) $\lim \limits_{\boldsymbol{V} \rightarrow \boldsymbol{c}} E=\quad a_1 h f=\quad \sqrt{1-\beta^2} /(1-n \boldsymbol{V} / \boldsymbol{c}) h f_0=\quad \sqrt{1-\beta^2} h f_0=$ $\sqrt{(Q-1) /(2 \ln Q)} h f_0$
and $\boldsymbol{p}=E / k \approx E$.

b) $\lim  \limits_{\boldsymbol{V} \rightarrow -\boldsymbol{c}} E=\quad a_1 h f=\sqrt{1-\beta^2} /(1+\quad n \boldsymbol{V} / \mathbf{c}) h f_0=\sqrt{1-\beta^2} h f_0=$ $\sqrt{(Q-1) /(2 \ln Q)} h f_0$
and $\boldsymbol{p}=E / k \approx E$.

In the Lorentz model, as we know that for the case where $\boldsymbol{V} \rightarrow \boldsymbol{c}, E$ tends to be infinite, however, the above we constrain $E$ to be a limited value that even if in the case where $\boldsymbol{V}=n c, f$ is infinite and $\lambda=0$, but $E=$ $\sqrt{1-\beta^2} /\left(1-\boldsymbol{v}_{\text {wave shock }}{ }^2 / c^2\right) h f_0$ is also a limited value. It is the limited energy and momentum that makes Compton scattering valid.

In addition, it has been stated above that due to $Q \approx 0$ there should be $a_1 \approx 1, a_2 \approx 1$ at low or medium frequency shift, which means that it should return to the case as in the Lorentz model at low or medium frequency shift.

Based on the three key considerations above, we construct the expression for $a_1$ and $a_2$ as Eq. (\ref{eq25}). One may consider other possible expressions for $a_1$ and $a_2$, but after many attempts we find that Eq. (\ref{eq25}) is the most concise expression that meets the above three requirements. Here one may be confused that why the correction coefficient corresponding to photon's energy and momentum are different, which seems to violate the intuition obtained from the Lorentz model. The main reason is that there is an invariant in the Lorentz model, which is the speed of light, but in our model, there is no such an invariant. However, although the correction coefficients for energy and momentum are different, they are still closely related (the underlying mechanism is that time and space are a whole, while Eq.(\ref{eq1}) has indicated that space and time are a whole), which is shown in Eq. (\ref{eq2}). So here in this paper we will just try to use this expression to analyze the phenomenon called synchrotron radiation in the cyclotron maser.

Now we have known that Eq. (\ref{eq23}) returns to Eq. (\ref{eq24}) at low or medium frequency shift. Thus next we will discuss the case where $\boldsymbol{v} \rightarrow \boldsymbol{c}$, and in this case we can obtain $\theta_2 \approx 1 / \gamma \approx 0$, then Eq. (\ref{eq23}) can be written accordingly as
\begin{equation}
    \label{eq28}
\begin{aligned}
\lambda_2= & \frac{n^{\prime}-\boldsymbol{v} / \boldsymbol{c}}{\sqrt{1-\beta^2}}\left[T_{c}  \frac{n}{a_2}\left(1+\frac{k}{n \boldsymbol{c}} \frac{\beta \cos \theta_1-\beta^2}{1-\beta \cos \theta_1}\right)\right]+\frac{h}{m_0} \sqrt{1-\beta^2}(n-\boldsymbol{v} / \boldsymbol{c}) \frac{1-\cos \phi}{\left(1-\beta \cos \theta_1\right)(1-\beta)}
\end{aligned}
\end{equation}
where $n^{\prime}=n\left(\boldsymbol{v} \cos \theta_2, \boldsymbol{c}\right) \approx n(\boldsymbol{v}, \boldsymbol{c})$, and
\begin{equation}
    \label{eq29}
a_2 \approx n \frac{1-\frac{\boldsymbol{v} \cos \left(\pi-\theta_1^{\prime}\right)}{n c}}{1-n \frac{\boldsymbol{v}}{\boldsymbol{c}} \cos \left(\pi-\theta_1^{\prime}\right)}=n \frac{1+\frac{\boldsymbol{v} \cos \theta_1^{\prime}}{n \boldsymbol{c}}}{1+n \frac{\boldsymbol{v}}{\boldsymbol{c}} \cos \theta_1^{\prime}}=n \frac{1+\frac{\boldsymbol{v}}{n \boldsymbol{c}} \frac{\cos \theta_1-\beta}{1-\beta \cos \theta_1}}{1+n \frac{\boldsymbol{v}}{\boldsymbol{c}} \frac{\cos \theta_1-\beta}{1-\beta \cos \theta_1}}
\end{equation}
When $\theta_1=0, \cos \theta_1{ }^{\prime}=(1-\beta) /(1-\beta)=1$, then it means $\theta_1{ }^{\prime}=0$, so $n=n$ $\left(\boldsymbol{v} \cos \left(\pi-\theta_1{ }^{\prime}\right), \boldsymbol{c}\right)=n(-\boldsymbol{v}, \boldsymbol{c})=n(\boldsymbol{v}, \boldsymbol{c})$ in Eq. (\ref{eq17}). Then we can obtain

\begin{equation}
\label{eq30}
\begin{aligned}
\lim _{v \rightarrow c} \lambda_2& =\lim _{v \rightarrow c} \frac{n-\boldsymbol{v} / \boldsymbol{c}}{\sqrt{1-\beta^2}}\left[T_c \frac{n(1+n \boldsymbol{v} / \boldsymbol{c})}{n+\boldsymbol{v} / \boldsymbol{c}}\left(1+\frac{1}{n} \beta\right)\right] \\
& +\frac{h}{m_0} \sqrt{1-\beta^2}(n-\boldsymbol{v} / \boldsymbol{c}) \frac{1-\cos \phi}{(1-\beta)^2} \\
& =\lim _{v \rightarrow c} \frac{T_c}{\sqrt{1-\beta^2}} \frac{n(n-\boldsymbol{v} / \boldsymbol{c})(1+n \boldsymbol{v} / \boldsymbol{c})}{n+\boldsymbol{v} / \boldsymbol{c}}+\frac{T_c}{\sqrt{1-\beta^2}} \frac{(n-\boldsymbol{v} / \boldsymbol{c})(1+n \boldsymbol{v} / \boldsymbol{c})}{n+\boldsymbol{v} / \boldsymbol{c}} \beta \\
& +\frac{h}{m_0} \sqrt{1-\beta^2}(n-\boldsymbol{v} / \boldsymbol{c}) \frac{1-\cos \phi}{(1-\beta)^2} \\
& \approx 0-T_c \sqrt{\frac{Q-1}{2 \ln Q}}-\frac{h}{m_0} \sqrt{\frac{Q-1}{\ln Q}} \frac{1-\cos \phi}{\left(1-\sqrt{1-\frac{Q-1}{\ln Q}}\right)^2} \\
& \approx-2 \pi R \sqrt{\frac{2 \ln Q}{Q-1}}
\end{aligned}
\end{equation}

In Eq. (\ref{eq30}) the following equation from Ref. \cite{b14} is used
\begin{equation}
    \label{eq31}
\begin{aligned}
\lim _{\boldsymbol{v} \rightarrow \boldsymbol{c} } \frac{1}{\gamma} & =\lim _{\boldsymbol{v} \rightarrow \boldsymbol{c}} \sqrt{1-\beta^2}=\lim _{\boldsymbol{v} \rightarrow \boldsymbol{c}} \sqrt{1-\left(\frac{\boldsymbol{v}}{k}\right)^2}=\lim _{\boldsymbol{v} \rightarrow \boldsymbol{c}} \sqrt{\frac{1-\boldsymbol{v} / \boldsymbol{c}}{n}(n+\boldsymbol{v} / \boldsymbol{c})} \\
& =\sqrt{\frac{Q-1}{2 \ln Q}}
\end{aligned}
\end{equation}
where the expression for $n$ is shown in Eq. (\ref{eq8}).

On the other hand, when $\theta_1=\pi, \cos \theta_1{ }^{\prime}=(-1-\beta) /(1+\beta)=-1$, then it means $\theta_1{ }^{\prime}=\pi$, so $n=n\left(v \cos \left(\pi-\theta_1{ }^{\prime}\right), \boldsymbol{c}\right)=n(\boldsymbol{v}, \boldsymbol{c})$ in Eq. (\ref{eq17}). Then we can obtain

\begin{equation}
\label{eq32}
\begin{aligned}
\lim _{\boldsymbol{v} \rightarrow \boldsymbol{c}} \lambda_2 & =\lim _{\boldsymbol{v} \rightarrow \boldsymbol{c}} \frac{n-\boldsymbol{v} / \boldsymbol{c}}{\sqrt{1-\beta^2}}\left[T_{c} \frac{n k / \boldsymbol{c}(1-n \boldsymbol{v} / \boldsymbol{c})}{n-\boldsymbol{v} / \boldsymbol{c}}\left(1-\frac{k}{n \boldsymbol{c}} \beta\right)\right] \\
& -\frac{h}{m_0} \sqrt{1-\beta^2}(n-\boldsymbol{v} / \boldsymbol{c}) \frac{1-\cos \phi}{(1-\beta)^2} \\
& =\lim _{\boldsymbol{v} \rightarrow \boldsymbol{c}} \frac{T_{\boldsymbol{c}}}{\sqrt{1-\beta^2}} n k / \boldsymbol{c}(1-n \boldsymbol{v} / \boldsymbol{c})-\frac{T_{c}}{\sqrt{1-\beta^2}} k^2 / \boldsymbol{c}^2(1-n \boldsymbol{v} / \boldsymbol{c}) \beta \\
& -\frac{h}{m_0} \sqrt{1-\beta^2}(n-\boldsymbol{v} / \boldsymbol{c}) \frac{1-\cos \phi}{(1-\beta)^2} \\
& =0-T_{c} \sqrt{\frac{2 \ln Q}{Q-1}}+\frac{h}{m_0} \sqrt{\frac{Q-1}{\ln Q}} \frac{1-\cos \phi}{\left(1-\sqrt{1-\frac{Q-1}{\ln Q}}\right)^2} \\
& \approx-2 \pi R \sqrt{\frac{2 \ln Q}{Q-1}}
\end{aligned}
\end{equation}

One may notice that the obtained value in Eqs. (\ref{eq30}) and (\ref{eq32}) are negative, which seems to go against our ``common sense". The reason is that, when $v>v_{\text {wave shock }}$, the timing sequence is reversed, which has been discussed in Section ``Doppler effect of photons". And it can be observed from Eqs. (\ref{eq30}) and (\ref{eq32}) that when $v=n, \lambda_2=0$ (here it should be noted that, although $\lambda_2=0$ in this case, both the momentum and energy of photons are finite, which has been stated in the paragraph following Eq. (\ref{eq27})), and when $\boldsymbol{v} \in(n, 1), \lambda_2$ increases dramatically and tends to be another non-zero value, which is obviously different from the results predicted by the Lorentz model that in the case where $\boldsymbol{v} \rightarrow \boldsymbol{c}$, the observed wavelength of light by the observer in the laboratory system is close to 0 (correspondingly, the momentum and energy of the photon are infinite).

\section{Inverse Compton scattering}\label{sec5}

The interaction between photons and other particles is often occurred in the high energy field, for example, the interaction between gamma-photons and nuclei, the interaction between gamma-photons and nucleons, etc. The scattering between extremely relativistic particles and photons is called inverse Compton scattering process, for that photons can obtain energy from particles in this process.

Unlike the Compton scattering process, in which the particles are basically stationary, in the inverse Compton scattering process the particles are extremely relativistic. However, their method of deriving the relationship between the scattered photon and the incident photon is the same, that is, they all use the law of energy conservation and momentum conservation shown in Eq. (\ref{eq12}). First we can build an inertial system $S^{\prime}$ that moves along with the electron (here we analyze the interaction between photons with electrons). In $S^{\prime}$, the electrons are stationary, so the collision between photons and electrons satisfies the Compton's formula as Eq. (\ref{eq18})

\begin{equation}
    \label{eq33}
\frac{\lambda_2^{\prime}}{a_2^{\prime}}-\frac{\lambda_1^{\prime}}{a_2} \approx \frac{h}{m_0}\left(1-\cos \phi^{\prime}\right)
\end{equation}
where $\lambda_1{ }^{\prime}$ and $\lambda_2{ }^{\prime}$ represent the wavelengths of incident and scattered photons observed in $S^{\prime}$ respectively, $\varphi^{\prime}$ denotes the scattering angle.

Here we assume that both the observer and the light source are located in the inertial system $S$, and that $S$ is moving at a velocity $v$ relative to $S^{\prime}$, then based on Eqs. (\ref{eq6}) and (\ref{eq25}) it has
\begin{equation}
    \label{eq34}
\begin{aligned}
& \lambda_1=n\left(1-\frac{\boldsymbol{v} \cos \theta}{n \boldsymbol{c}}\right) \frac{\lambda_1^{\prime}}{\sqrt{1-\beta^2}} \\
& a_2=n \frac{1-\frac{\boldsymbol{v} \cos \theta}{n \boldsymbol{c}}}{1-n \frac{\boldsymbol{v} \cos \theta}{\boldsymbol{c}}} \\
& \lambda_2=n\left(1-\frac{\boldsymbol{v} \cos \theta}{n \boldsymbol{c}}\right) \frac{\lambda_2^{\prime}}{\sqrt{1-\beta^2}}
\end{aligned}
\end{equation}
where $\lambda_1$ and $\lambda_2$ represent the wavelengths of incident and scattered photons observed in $S$ respectively.

Next we consider the case where $\theta=0$ (for that in this case a maximum energy can be obtained by the photon from the particle), and correspondingly it has $\varphi^{\prime}=\pi$. Since Eq. (\ref{eq33}) describes the Compton scattering that occurs in $S^{\prime}$, we can assume that $a_2{ }^{\prime} \approx 1$.

Then based on Eqs. (\ref{eq33}) and (\ref{eq34}), the observer in $S$ can obtain the scattered photons with momentum $\boldsymbol{P}_2$

\begin{equation}
    \label{eq35}
\begin{aligned}
\boldsymbol{P}_2& =a_2^{\prime \prime} \frac{h}{\lambda_2}=\frac{\boldsymbol{n}-\boldsymbol{v} / \boldsymbol{c}}{1-\boldsymbol{nv} / \boldsymbol{c}} \frac{h}{\frac{n-\boldsymbol{v} / \boldsymbol{c}}{\sqrt{1-\beta^2}}\left[\frac{2 h}{m_0}+(1-n \boldsymbol{v} / \boldsymbol{c}) \frac{\lambda_1}{\sqrt{1-\beta^2}}\right]} \\
& =\frac{1}{1-n \boldsymbol{v} / \boldsymbol{c}} \frac{h}{\frac{1}{\sqrt{1-\beta^2}} \frac{2 h}{m_0}+(1-n \boldsymbol{v} / \boldsymbol{c}) \frac{\lambda_1}{1-\beta^2}}
\end{aligned}
\end{equation}

When $n \equiv 1$, Eq. (\ref{eq35}) returns to the case as in the Lorentz model. From Eq. (\ref{eq35}) it can be seen that when $n=v / c, \boldsymbol{P}_2$ has a maximum value, that is

\begin{equation}\label{eq36}
\begin{aligned}
&\begin{aligned}
\boldsymbol{P}_2^{\max } & =\frac{1}{1-n \boldsymbol{v} / \boldsymbol{c}} \frac{h}{\frac{1}{\sqrt{\frac{1-\boldsymbol{v} / \boldsymbol{c}}{n}(n+\boldsymbol{v} / \boldsymbol{c} )}} \frac{2 h}{m_0}+(1-n \boldsymbol{v} / \boldsymbol{c})_{\frac{\lambda_1}{\frac{1-\boldsymbol{v} / \boldsymbol{c}}{n}(n+\boldsymbol{v} / \boldsymbol{c})}}} \\
& \approx \frac{h}{\frac{4 \sqrt{2} h}{m_0} \sqrt{1-v_{\text {wave shock }}}+\lambda_1\left(1-v_{\text {wave shock }}\right)}
\end{aligned}
\end{aligned}
\end{equation}

where $v_{\text {wave shock }}$ comes from Eq. (\ref{eq9}).

For the observer in $S$, if the observed energy of the electron is much greater than that of the incident photon, Eq. (\ref{eq36}) can be further simplified as

\begin{equation}
\label{eq37}
\boldsymbol{P}_2^{\max } \approx \frac{\sqrt{2} m_0}{8 \sqrt{1-v_{\text {waveshock }}}}
\end{equation}

And in addition, what we are more interested in is that what would happen if $\boldsymbol{v} \rightarrow \boldsymbol{c}$, in which case $\boldsymbol{n} \rightarrow 0$. Again, based on Eq. (\ref{eq35}) we can obtain

\begin{equation}
\label{eq38}
\lim _{\boldsymbol{v} \rightarrow \boldsymbol{c}} \boldsymbol{P}_2=\lim _{\boldsymbol{v} \rightarrow \boldsymbol{c}} \frac{h}{\frac{1}{\sqrt{1-\beta^2}}\frac{2 h}{m_0}+\frac{\lambda_1}{1-\beta^2}}=\frac{h}{\sqrt{\frac{2 \ln Q}{Q-1}} \frac{2 h}{m_0}+\frac{2 \ln Q}{Q-1} \lambda_1}
\end{equation}

As one knows, in the Lorentz model, when the velocity of the electron is close to the speed of light, the maximum energy that a photon can obtain through inverse Compton scattering process tends to be infinite. However, if $Q$ is not equal to 0 (when $Q=0$, then $n \equiv 1$ and the model discussed above will return to the Lorentz model), then the maximum energy that a photon can obtain through inverse Compton scattering process will have a limit, which is shown in Eq. (\ref{eq37}), and further when the velocity of electron is high enough to be close to $c$, the maximum energy that a photon can obtain instead drops to another non-zero and limited value.

In addition, from Eq. (\ref{eq37}) it can be seen that the maximum energy that a photon can obtain is not related to the wavelength of the incident photon, and the ratio of the maximum energy that a photon can obtain to the rest energy of an electron is a constant, which is very interesting. That is, in inverse Compton scattering, no matter what the energy of the incident photon is, and no matter whether the extremely relativistic particle is an electron or a proton or other particle with mass, the ratio of the maximum energy that a photon can obtain to the rest energy of the extremely relativistic particle is a constant.

In recent years, the mechanism of high energy radiation from blazars have been studied extensively \cite{b25,b26,b27,b28,b29,b30,b31}. One model used to explain this mechanism is the lepton origin model, which considered that the high energy radiation comes from the inverse Compton scattering process between relativistic electrons and low energy photons \cite{b29,b30,b31}. Since some special objects with extreme physical properties, such as the blazars, can produce extremely relativistic particle jets, and the microwave background radiation (CMB) provides natural low-energy photons, then there provides a cosmic version of inverse Compton scattering. Therefor, we expect to test the value of $Q$ by analyzing the ultra-high energy radiation spectrum in such a cosmic scenario, for that whether the value of $Q$ is 0 will have a very different effect on the ultrahigh energy radiation spectrum, which is shown in this section.

\section{Summary}\label{sec6}

The Lorentz violation model is widely studied in the ultra-high energy field in recent years, especially is considered in various models attempting to unify the general relativity and quantum theory \cite{b32,b33,b34,b35,b36,b37}. Among them Ref. \cite{b14} proposed a possible Lorentz violation model that the speed of light may be relate to the velocity of light source. The key idea in Ref. \cite{b14} is that in the case where $\boldsymbol{v} \rightarrow \boldsymbol{c}$ the energy or momentum of the particle have a limited value (i.e., the limit of the time or length scaling factor is $\sqrt{2 \ln Q /(Q-1)}$, and correspondingly, the energy limit of a particle with rest mass $m_0$ is $m_0 c^2 \sqrt{2 \ln Q /(Q-1)} /[1-0.5(Q-1) /$ $\ln Q])$ rather than be infinite predicted by the Lorentz model. The model in Ref. \cite{b14} is similar to the famous rainbow model \cite{b12}, but they are different, for that in Ref. \cite{b14} the energy limit of the particle is related to the rest mass of the particle, while in the rainbow model it is not, but is assumed to be a constant (the constant is usually considered to be the Planck energy or near the Planck energy). If we follow the idea in Ref. \cite{b14} that the speed of light observed by an observer moving with a velocity $\boldsymbol{v}$ relative to the light source is $n \boldsymbol{c}$, then it will forces us to have to modify the usual quantum relation of photons in the Doppler effect.

By discussing the Doppler effect of photons based on the idea in Ref. \cite{b14}, we find that light can also arise the shock phenomena (in vacuum) just like sound wave, and further we obtain the conditions for light to generate a shock wave, that is, $\boldsymbol{v}=n \boldsymbol{c}$, and in the case where $n \boldsymbol{c}<\boldsymbol{v}<\boldsymbol{c}$ the timing sequence will be reversed, but it doesn't violate the law of causality for that the timing sequence can be reversed in independent events (but not in dependent events). Then the Compton scattering process between the photon with the modified quantum relation as shown in Eq. (\ref{eq11}) and the electron is discussed, and we find that unless we define a suitable correction coefficient for the quantum relation of photons, the photons energy will be infinite in the case where $\boldsymbol{v}=n \boldsymbol{c}$, which makes it impossible to discuss the Compton scattering process  between photons and electrons over the whole range of $\boldsymbol{v} \in[0, \boldsymbol{c})$, in other words, due to the infinite energy of photon in the case where $\boldsymbol{v}=$ $n \boldsymbol{c}$, the Compton scattering between photons and electrons will be invalid. So in order to avoid that we defined a suitable and concise correction coefficient for the quantum relation of photons to make the energy of photons have a limited value over the whole range of $\boldsymbol{v} \in[0$, $\boldsymbol{c})$, especially in the case where $\boldsymbol{v}=n \boldsymbol{c}$ or $\boldsymbol{v} \rightarrow \boldsymbol{c}$.

Here it should be noted that, due to the limited information or knowledge we have at present, we are not yet able to obtain the specific correction coefficients for the quantum relation of photon satisfying Eq. (\ref{eq2}), but we can obtain three key constraints for the possible correction coefficients, which is shown in Section ``Spontaneous radiation in cyclotron maser". One may propose a lot of expressions for the correction coefficients satisfying the three constraints, but here in this paper after many attempts we construct a very concise expression for the correction coefficients, which is shown in Eq. (\ref{eq25}).

Then based on Eq. (\ref{eq25}) we choose the phenomenon called spontaneous radiation in a cyclotron maser to study, for that the phenomenon can be explained by the concept of virtual photon. By analyzing the Doppler effect and Compton scattering effect of virtual or real photons in the cyclotron maser, we can obtain the wavelength formulas of spontaneous radiation and synchrotron radiation. The important conclusion we have drawn is that we find in the case where $\boldsymbol{v} \rightarrow \boldsymbol{c}$ the wavelength observed by the observer in the laboratory system does not tend to 0 , but tends to a non-zero value, which is different from the results derived from the Lorentz model that the wavelength tends to 0 and the corresponding energy and momentum of photon tend to be infinite. Even in the case where $\boldsymbol{v}=n \boldsymbol{c}$, the wavelength is 0 , but the energy and momentum of photon are both finite, which ensures the Compton scattering is valid over the whole range of $\boldsymbol{v} \in[0, \boldsymbol{c})$, since in Ref. [14] it allows $\boldsymbol{v} \in$ $[0, \boldsymbol{c})$

Further, the inverse Compton scattering phenomenon based on Eq. (\ref{eq25}) is also discussed, and we find that there is a limit to the maximum energy that can be obtained by photons in the collision process between relativistic particles and low-energy photons, and this case occurs when the velocity of particle satisfies $v=v_{\text {wave shock }}$. When the velocity of particle is greater than $\boldsymbol{v}_{\text {wave shock }}$ and close to $\boldsymbol{c}$, the energy that can be obtained by the photon decreases sharply and tends to be another limited value. This conclusion is also very different from that predicted by the Lorentz model, in which the energy that can be obtained by the photon tends to be infinite as the velocity of particle approaches to $\boldsymbol{c}$.

Generally, the main purpose of this paper is the same as that of Ref. \cite{b14}, that is, to avoid the energy or momentum of particles (i.e., any particles, including photons) to be infinite (otherwise it will make the real physical situation invalid in certain cases), and which is also inspired by the idea in some quantum gravity models \cite{b34,b35,b37}. It is very important to emphasize that if $Q=0$, then $n \equiv 1$ and correspondingly, all the results and conclusions in this paper will return to the case as in the Lorentz model. Therefor, this paper also show us another possible experimental scheme to determine the value of $Q$, although it still requires extremely high experimental energy.

\section*{CRediT authorship contribution statement}

\textbf{ Jinwen Hu: }Methodology, Investigation, Formal analysis, Conceptualization. \textbf{Huan Hu: } Data curation, Writing - original draft.

\section*{Declaration of competing interest}

The authors declare that they have no known competing financial  interests or personal relationships that could have appeared to influence  the work reported in this paper.

\section*{Data availability}
Data will be made available on request.

\bibliographystyle{elsarticle-num-names}
\bibliography{ref}

\end{document}